\documentclass{aastex63}
\usepackage[utf8]{inputenc}
\usepackage{graphicx} 
\usepackage{amsmath}
\usepackage{appendix}

\begin{document}

\newcommand{\two}{$N\hspace{-0.2em}\ge\hspace{-0.2em}2$ }
\newcommand{\three}{$N\hspace{-0.2em}\ge\hspace{-0.2em}3$ }
\newcommand{\four}{$N\hspace{-0.2em}\ge\hspace{-0.2em}4$ }

\title{Architectures of Planetary Systems II: Trends with Host Star Mass and Metallicity}

\author[0000-0002-4884-7150]{Alex R. Howe}
\affiliation{The Catholic University of America, 620 Michigan Ave., N.E. Washington, DC 20064}
\affiliation{NASA Goddard Space Flight Center, 8800 Greenbelt Rd, Greenbelt, MD 20771, USA}
\affiliation{Center for Research and Exploration in Space Science and Technology, NASA/GSFC, Greenbelt, MD 20771}

\author[0000-0002-7733-4522]{Juliette C. Becker}
\affiliation{Department of Astronomy, University of Wisconsin-Madison, 475 N. Charter Street, Madison, WI 53706, US}


\author{Fred C. Adams}
\affiliation{Department of Physics, University of Michigan, 450 Church St, Ann Arbor, MI 48109}
\affiliation{Department of Astronomy, University of Michigan, Ann Arbor, MI 48109}

\begin{abstract}
The current census of planetary systems displays a wide range of architectures. Extending earlier work, this paper investigates the correlation between our classification framework for these architectures and host stellar properties. Specifically, we explore how planetary system properties depend on stellar mass and stellar metallicity. This work confirms previously detected trends that jovian planets are less prevalent for low-mass and low-metallicity stars. We also find new, but expected trends such as that the total mass in planets increases with stellar mass, and that observed planetary system masses show an upper limit that is roughly consistent with expectations from the stability of circumstellar disks. We tentatively identify potential unique trends in the host stars of super-puffs and hot jupiters and a possible subdivision of the class of hot jupiter systems. In general, we find that system architectures are not overly dependent on host star properties.
\end{abstract}

\section{Introduction}
\label{sec:intro}

With over 6000 confirmed planets discovered outside our Solar System, the database has reached a point where it is useful to provide a classification for planetary systems as astrophysical objects in themselves. In our previous paper, \cite{PaperI}, hereafter Paper I, we introduced a new framework for classifying planetary systems based on their dynamical architectures. This work was the first such classification framework to include the entire catalog of confirmed exoplanets, considering both the masses and orbital periods of the planets. In this paper, we extend this analysis of planetary systems to include the properties of their host stars.

\begin{figure}[!ht]
    \centering
    \includegraphics[width=0.66\textwidth]{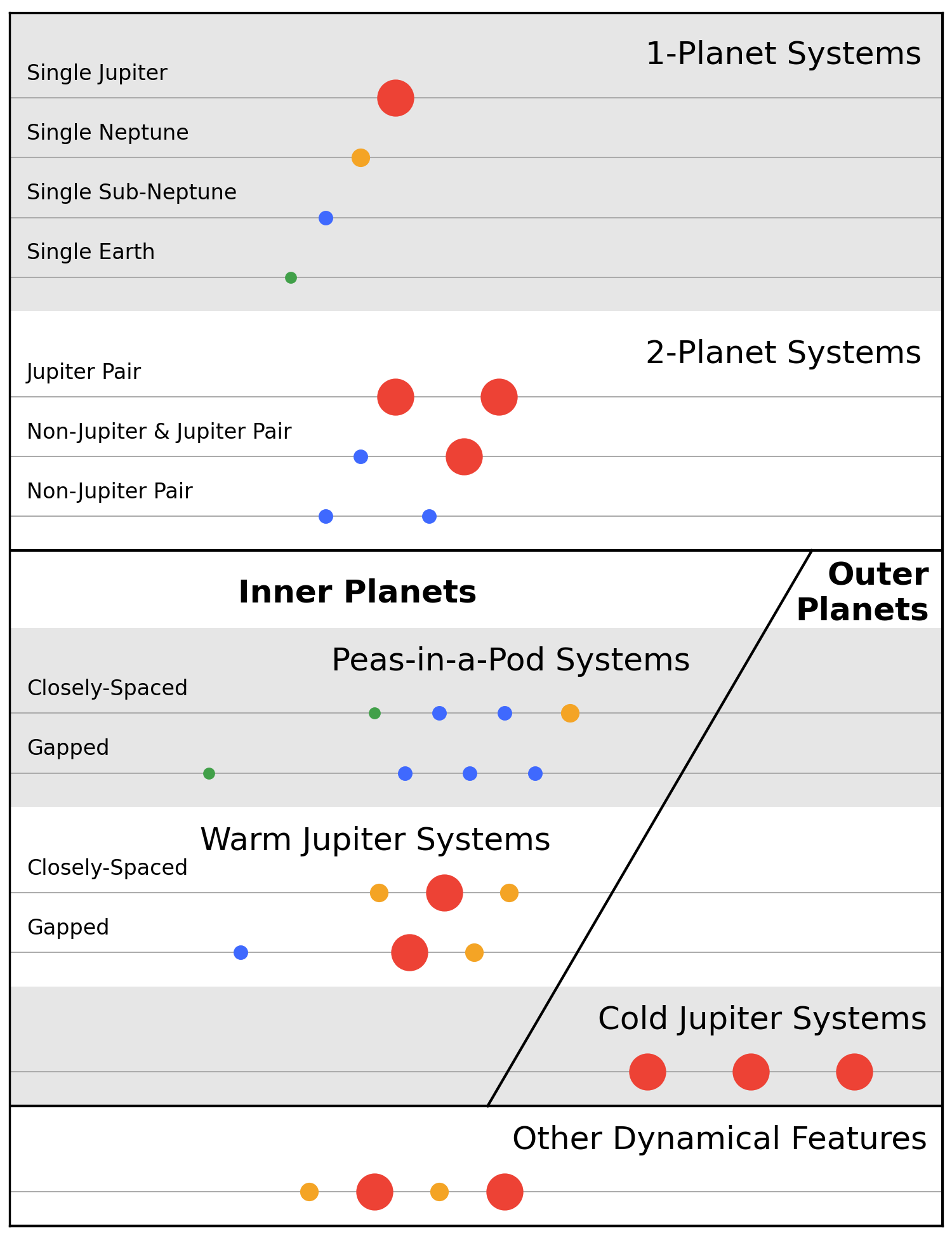}
    \caption{Abridged quick-reference chart for our classification of planetary system architectures presented in Paper I. Each row corresponds to one planetary system, with horizontal spacing corresponding to orbital period on a log scale and point sizes corresponding to planet size. Colors correspond to planet type: jupiters in red, neptunes in yellow, sub-neptunes in blue, and earths in green.}
    \label{fig:quick_reference}
\end{figure}

We provide an abbreviated schematic for our classification framework in Figure \ref{fig:quick_reference}. The core of this classification comes down to three questions for any given system. Does the system have distinct inner and outer planets? Do the inner planets include one or more Jupiters? And do the inner planets contain any gaps with a period ratio greater than 5? We find that these three questions are sufficient to classify $\sim$97\% of multiplanet systems with \three planets with minimal ambiguity. In particular, this framework allows us to neatly divide the systems with \three inner planets into four categories: the closely-spaced peas-in-a-pod systems (hereafter CPP systems), the gapped peas-in-a-pod systems (GPP), the closely-spaced warm jupiter systems (CWJ), and the gapped warm jupiter systems (GWJ). Hot jupiters (HJ) can also be considered as their own class, albeit with a small amount of overlap.

The details of our classification system and a catalog of the categories of \three systems are provided in Paper I. We note that in the full classification scheme, we add additional subcategories of ``gapped'' systems based on where in the system the gap occurs. We do not address these subcategories in this paper because of their small sample sizes. We also identify other dynamical features of interest in Paper I, i.e., systems that are dynamically unusual in ways not captured by our general framework. We investigate the host star populations for these categories in Section \ref{sec:special}.

While our classification framework focuses on the structures of planetary systems, it can provide additional insight into other properties of these systems, such as the characteristics of their host stars, based on how and whether physical properties such as host star mass and metallicity correlate with these structural categories. Any such correlations could point to star-dependent differences in evolutionary processes and as such could provide additional guidance for target selection in the search for potentially habitable planets.

In this paper, we perform a statistical analysis (see Section \ref{sec:methods}) of the host star populations for the categories of multiplanet systems in our classification to investigate whether they are consistent with broader potential source populations, such as stars with or without jupiters. While the signal sizes are generally small, the stellar properties appear to be correlated with the total planet mass in each system, but not otherwise with their dynamical properties, with few exceptions (see Sections \ref{sec:hj} and \ref{sec:sp}). 

This statistical analysis is challenging both because of the large selection effects in the detected exoplanet population and because of the small sample sizes of many of the categories. The former concern is partially mitigated by the fact that, throughout this paper, we compare one set of observed exoplanets to another, so the selection effects on each set should be similar. In the latter case, the small sample sizes necessarily limit the scope of our analysis, and the lower completion rate of long-period planets for the most part restricts us to searching for correlations with short-period planets. Nonetheless, our analysis can shed light on the parts of the parameter space that \textit{are} readily detectable and well-characterized. This means that we have found most of the mass in short-period planets in each system, and even this subset of the parameter space will provide useful insights into planet formation and will help to identify statistically significant trends more broadly.

For most of the dynamical categories of systems considered herein, we generally find that their host stars are derived from the ``expected'' source populations of jupiter-hosting or jupiter-free stars, as appropriate. While the signal sizes in our results are generally small, we are able to rule out large deviations from the expected patterns, and this analysis proves to be sufficient to identify trends in the host star population with total planet mass, as well as the tentative anomalous results in the cases of super-puffs and hot jupiters.

This paper is organized as follows. We begin by updating our census of planets in Section \ref{sec:census} and summarizing the properties of the exoplanet host star population in Section \ref{sec:stars}. We describe our statistical methods in Section \ref{sec:methods} and the results of our statistical analyses for stellar mass and metallicity in Section \ref{sec:results}. Finally, we discuss the implications of our results in Section \ref{sec:discussion}.

\section{Updated Dataset and Filtering}
\label{sec:census}

As in Paper I, our data (for both stars and planets) are drawn from the Planetary Systems Composite Parameters Table \citep{Archive} of the NASA Exoplanet Archive \citep{Christiansen2025}, in this case, specifically, the October 2, 2025 version. This version of the Archive contains a net 254 more planets than the version used in Paper I, the majority discovered by the \textit{TESS} spacecraft and the rest by other surveys including transits, radial velocities, microlensing, and astrometric observations from the \textit{Gaia} spacecraft. 

Due to missing, erroneous, or otherwise-unsuitable data in the Archive, we filtered the database for compatibility with our methods, recalculating some masses and radii based on the mass-radius relation of \cite{ChenKipping}, and in a few cases applying manual corrections based on the literature. Our methods for filtering the data and our list of manual corrections are detailed in Paper I. We make one additional manual correction here: for Kepler-80g, we adopt the 1.0 $M_\oplus$ mass from \cite{Kepler80ref}.

In addition to these changes, we have added one new filter for our analysis in this paper. Paper I was inconsistent in removing pre-Main Sequence stars (which do not represent a ``mature'' planet population) from the database, instead focusing on systems that are clear outliers from the population as a whole, such as V1298 Tau. In this paper, we apply this standard more rigorously and remove all stars from the dataset that are younger than 30 Myr, consistent with the timescale of planet formation, if their ages are listed in the catalog. 
This filter excludes 49 planets from our dataset that were included in Paper I, but these new exclusions do not have a significant effect on our classification framework. In particular, only one system with \three planets is removed: AU Mic. 


After accounting for new planets and changes in filtering (but before filtering for availability of stellar properties), the final number of planets in our adopted data set is 5881, increased from 5686 in Paper I. The number of planetary systems in our adopted data set is 4289, increased from 4259. Of these, 988 are multiplanet systems\footnote{Notably, in addition to the total number of detected exoplanets surpassing 6000, the total number of detected \textit{multiplanet} systems has now surpassed 1000.}, and 325 have at least three planets, enough for further classification. 

\section{Distribution of Host Star Properties}
\label{sec:stars}

The properties of host stars present a more complex problem for our analysis because the stars themselves are not always well-characterized, even when their planets are. Our filtering process requires that all planets in our sample have a known host star mass in order for semi-major axes and periods to be uniformly inter-converted. Such completeness is not needed \textit{a priori} for other stellar parameters such as luminosity (listed for 95\% of planets), effective temperature (95\%), metallicity (91\%), age (78\%), or spectral type (36\%). 

It is expected that host star mass \citep{Laughlin2004} and metallicity \citep{Fischer2005} should have the greatest impact on planet formation in general and giant planet formation in particular \citep[see also][and subsequent references]{Johnson2010occur}. 
Thus, we use stellar mass and metallicity as our primary variables to study the host star population. With our dataset listing metallicities for $>$90\% of host stars, we are able to thus characterize the large majority of the population.

To best highlight the correlation between stellar properties and planet formation, in this section, we have studied the total mass of all (known) planets in each system. This is a better proxy and should capture the properties of the inner disk at the time of formation more faithfully than considering individual planets. That is, we expect that any correlations between formation processes in the inner part of the disk with modern-day system architectures could plausibly leave detectable signals in the observed population, which probes these separations.

\begin{figure}[!ht]
    \centering
    \includegraphics[width=0.99\textwidth]{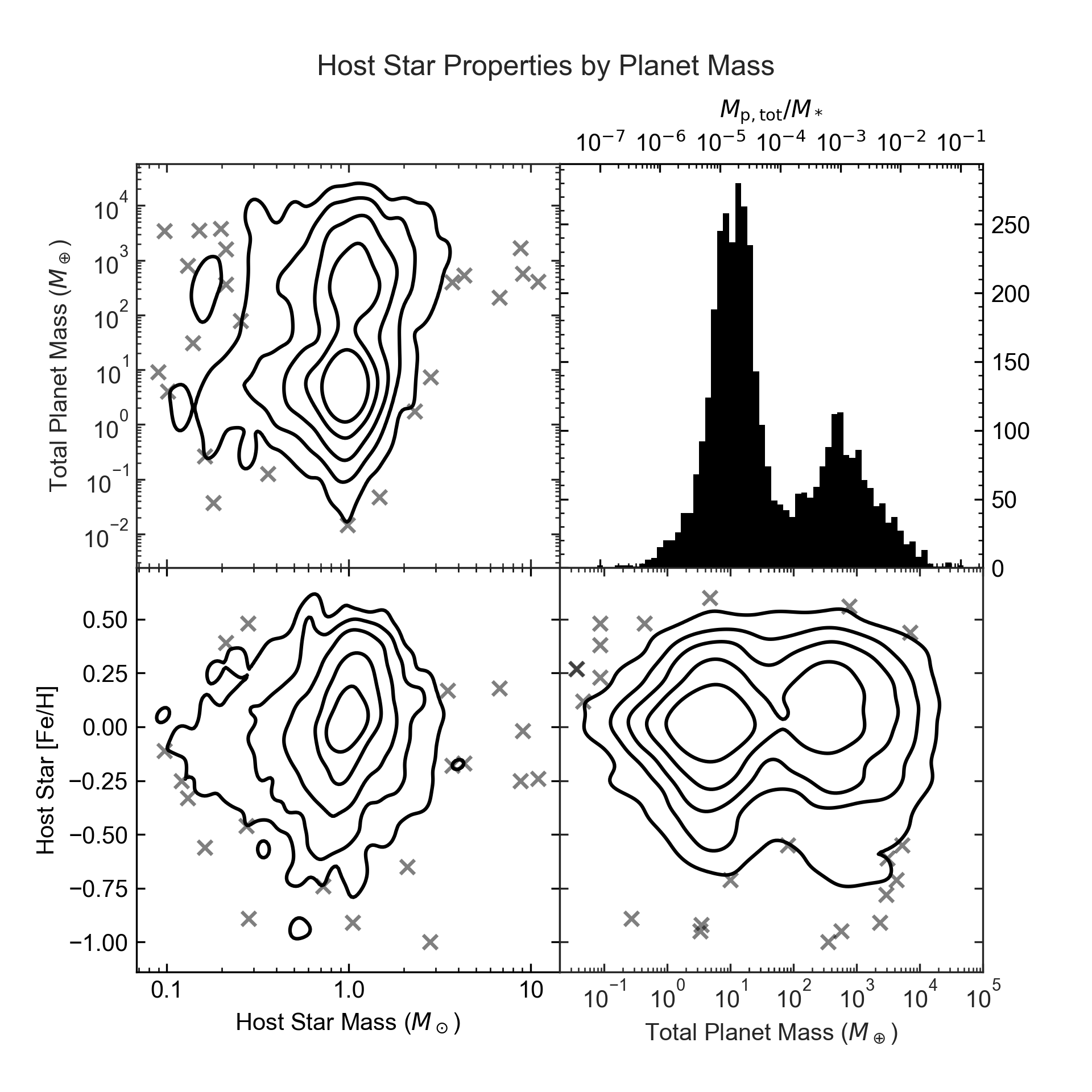}
    \caption{Masses and metallicities of the exoplanet host star population compared with the total planet mass in each system. Top left, bottom left, and bottom right are density contour plots where each contour corresponds to 0.5$\sigma$, with the outermost contour being 2.5$\sigma$, or 99\% of host stars. Outlier systems are marked with X's. Top right is a histogram of mass ratios of total planet mass divided by host mass.}
    \label{fig:starprop}
\end{figure}

The properties of the host star population compared with the respective masses of their planets are plotted in Figure \ref{fig:starprop}. Three of the panels of Figure \ref{fig:starprop} are density contour plots, where each contour represents $0.5\sigma$, and outliers beyond the $2.5\sigma$ contour (roughly the 99th percentile) are marked with X's. These subplots include host star mass versus metallicity (bottom left), total planet mass versus host mass (top right), and total planet mass versus host metallicity (bottom right). We also include a histogram (upper right) of mass ratios of total planet mass divided by host mass. We note that the total planet mass is bimodal, with one peak in the super-Earth range of $1-10\ M_\oplus$ and one peak near 1 $M_J$, reflecting the bimodality of detected planet masses generally. Because roughly three-fourths of detected planets are in single planet systems, these plots will look similar to equivalent plots for individual planet masses.

The largest mass ratios in our sample are approximately 0.03. Several systems lie near this value, though only one is both a non-directly-imaged system and unambiguously constrained not to include a brown dwarf: GJ 676A, which consists of two super-jupiters and two sub-neptunes orbiting a 0.63 $M_\odot$ M-dwarf. As the largest planets are also the most detectable, this result should be independent of selection effects (with the exception of widely separated planets at $>$100 AU, which may have different formation processes).

We also note that while correlations between the quantities being compared are quite weak, there is a noticeable correlation between host mass and planet mass and a slight correlation between host metallicity and planet mass. These findings suggest that both larger and more metal-rich stars produce larger planets, on average.

While these correlations are notable and are also consistent with our understanding of planet formation, they do not directly address correlations between host star properties and our classification framework. Our framework does not consider total planet mass under most circumstances, nor does it consider individual planet masses as a continuous distribution, but instead divides them into discrete size bins based on a modified version of the classification scheme of \citealt{Kopparapu18}.

\begin{figure}[!ht]
    \centering
    \includegraphics[width=0.99\textwidth]{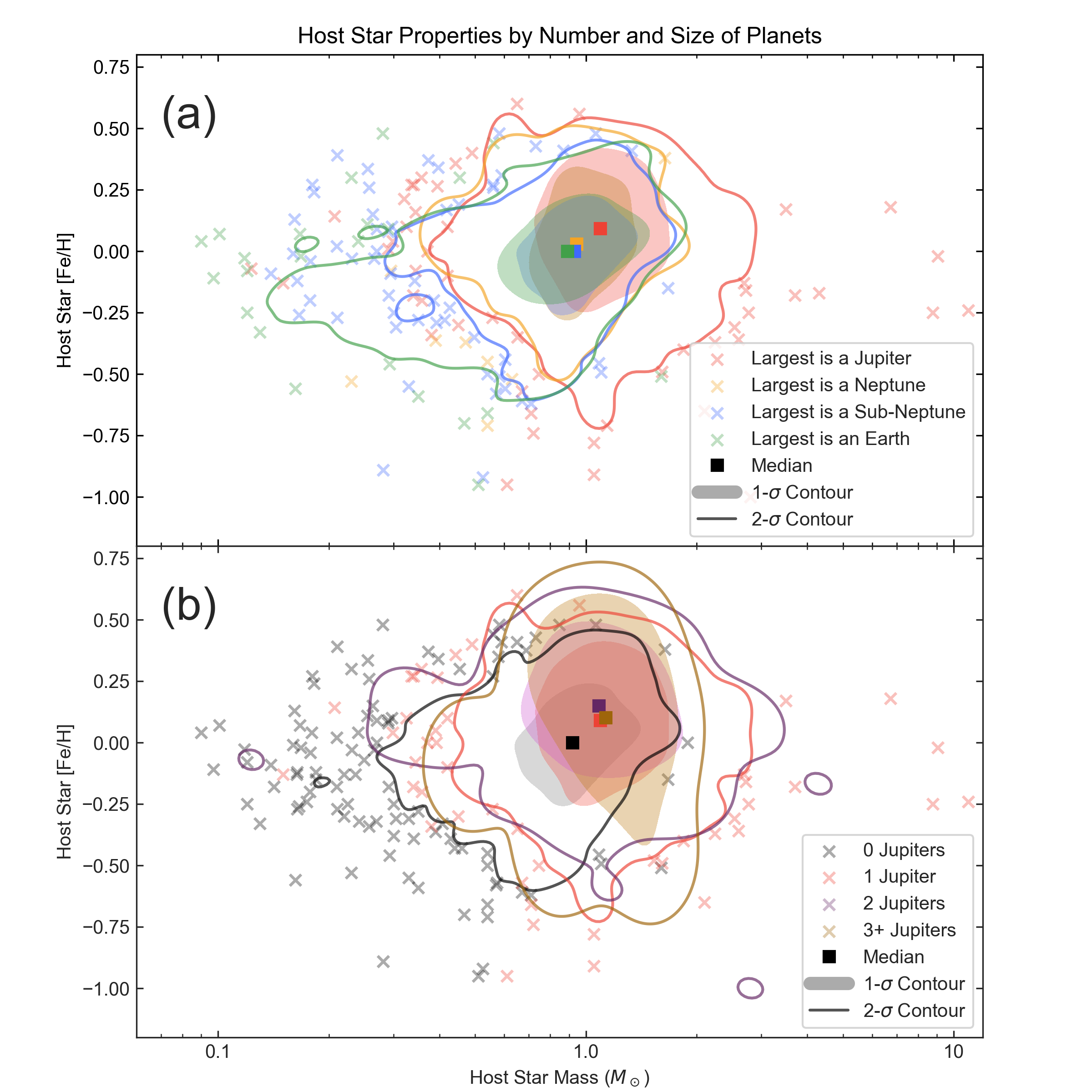}
    \caption{Masses and metallicities of exoplanet host stars divided by planet size and number. Square markers indicate the median points of each distribution. Shaded areas are $1\sigma$ density contours; solid lines are $2\sigma$ contours, and X's are $>2\sigma$ outliers. Panel (a): population divided into subsets based on whether the largest detected planet is a jupiter (red), a neptune (gold), a sub-neptune (blue), or an earth (green). Panel (b): population divided into subsets with no jupiters (black), exactly one jupiter (red), exactly two jupiters (purple), and three or more jupiters (dark yellow). (The highest number in our data set is four.)}
    \label{fig:nsize}
\end{figure}

To illustrate the mass dependence for individual planets, we draw density contour plots of host masses and metallicities for our four bins in planet mass in panel (a) of Figure \ref{fig:nsize}. Because our classification does not consider total planet mass, but \textit{does} consider the presence or absence of Jupiters important, we divide the bins by the size of the \textit{largest} known planet in each system. Systems where the largest planet is a jupiter are drawn in red, systems where the largest planet is a neptune in yellow, systems where the largest planet is a sub-neptune in blue, and systems with only earths in green. (As in Paper I, for planets whose masses are not known, their masses are computed from their measured radii based on the mass-radius relation of \citealt{ChenKipping}.) In panel (b) we extend this division by planet mass based on the \textit{number} of jupiters in each system: no jupiters in black, exactly one jupiter in red, exactly two jupiters in purple, and three or more jupiters in dark yellow (the maximum number in our dataset being four).

In this figure, to better illustrate the differences between populations, we limit the contours drawn to $1\sigma$ (shaded areas), $2\sigma$ (solid lines), and median points (squares), while outliers (now approximately the 95th percentile) remain marked with X's. (This format also applies to the density contour plots in Section \ref{sec:results}.)

These populations of host stars are broadly similar to one another in the parameter space. In terms of median values, the median host mass and metallicity for the jupiter-hosting population are both significantly higher than those of the other populations. However, the most pronounced differences appear in the tails of the distributions. The tail of the jupiter-hosting population extends to higher host masses and includes virtually all high-mass host stars $>2\ M_\odot$. Meanwhile, the earths-only population includes a much larger tail extending to small host masses of $<0.3\ M_\odot$. The neptune- and sub-neptune-hosting populations have only outlier systems in these tail regions, or none at all. Likewise, there is not a significant difference in the median host masses for systems with one jupiter versus more than one, but the tails of these distributions show significant differences between different numbers of jupiters.

Similar results hold in metallicity space. Specifically, the tail of the neptune-hosting population extends to higher metallicities than the sub-neptune and earths-only populations. Moreover, the tail of the jupiter-hosting population extends to still higher metallicities, increasing with the number of jupiters. 

Other than differences with the number of jupiters, we do not see strong trends in the host star population with system multiplicity. 
Still, the difference between jupiters and non-jupiters appears to be a robust result, even though it is not an across-the-board trend. This is reinforced by several of our statistical tests (described in Section \ref{sec:results}) showing significant differences between jupiter-bearing and jupiter-free hosts as source populations.

\section{Statistical Methods}
\label{sec:methods}

While we define many categories of planetary systems in Paper I based on their architectures, three-quarters of observed systems are single-planet systems (more precisely, systems with only a single known planet). This presents a unique problem when considering stellar properties as a function of system architectures because their completeness is likely to be low, with many undetected planets. The better-defined classes of \three systems are more promising, as a large majority of them are dynamically packed, so that we can be confident that they are complete (or nearly complete) for short-period planets. Additionally, the population of these dynamically packed systems in aggregate indicates that most of them are complete with respect to their outer edges, without additional planets in the ''pea pod'' at larger separations \citep{Millholland2019}. Thus \three systems should be a more reliable probe of inner systems. The limitation of these systems, however, is sample size; they are far fewer in number ($\sim$300 in total, the large majority of which fall in one class, CPP), making it much harder to draw conclusions about their populations.

Out of a range of possible comparisons, we find that the most informative method of analysis is to determine whether the stellar hosts for the various better-defined classes of \three systems can be consistently drawn from any of the larger potential source populations of stellar hosts, and how this varies among the classes compared with each other. 
Comparing populations of observed exoplanet systems with each other as opposed to e.g. field stars will mitigate many of the problems of completeness of observations and undetected planets because the sample of interest and the potential source populations will be subject to similar observational biases.

To simplify the space of possible source populations, we limit most our analyses to the four most clearly distinct populations: host stars with jupiters versus host stars without jupiters, and all host stars versus multiplanet hosts. This approach still poses problems for completeness, as outer jupiters especially are less likely to be detected with current methods, but they should be a good probe of the inner planets that are the primary focus of our investigation, especially because larger planets are less likely to go undetected than smaller ones. 

Nonetheless, comparing samples of interest with potential source populations remains a challenge in this context because of the small sizes of most of these classes. With so few data points, it will not be apparent if they are consistent with a given source population, or whether any apparent deviations are a result of random chance. Thus, a more rigorous statistical measure of the differences between the populations is needed.

A natural statistical measure for the consistency or inconsistency of two samples of data points is the $k$-nearest neighbors statistic. This statistic (or family of statistics) is a measure of how much the data points in each of two (or more) samples are clustered in parameter space with their own sample as opposed to the other. As used in this paper, the $k$-nearest neighbors statistic $T_{n,k}$ on a set of points $p$, consisting of two samples of combined size $n$, is given by \citep{statsbook}:
\begin{align}
    T_{n,k}&=\frac{1}{nk}\sum_{i=1}^n \sum_{j=1}^kI_i(j), \\
    I_i(j)&=
    \begin{cases}
       1,\ p_i \textrm{ and } p_j \textrm{ belong to the same sample.} \\ \nonumber
       0,\ p_i \textrm{ and } p_j \textrm{ belong to different samples.} \nonumber
    \end{cases} \\ \nonumber
\end{align}

This statistic computes the average fraction of each data point's $k$ nearest neighbors in the parameter space (in this case host mass and metallicity) that come from the same sample. (All neighbors are given equal weight for this purpose regardless of their ranking, and overlaps are removed to prevent double-counting.) If that fraction is higher than random chance; that is, if nearby points are more likely to be from the same sample than we would expect of two random samples from the same population would suggest, then the two samples can be said to be drawn from different parent populations.

In a Bayesian framework, the evidence for the null hypothesis (that the samples are drawn from the same) population can be computed based on the prior probability distributions of the populations. 
However, in this case, with large systematic uncertainties and the true population distribution being unknown, this is not possible, so we instead use a permutation test to obtain an empirical reference distribution for the $k$-nearest neighbors statistic. 

A permutation test allows hypothesis testing by comparing sample distributions in situations such as this one, where the prior probabilities are unknown or not meaningful, by resampling the available data. In this case, the samples being compared are the subpopulation of interest and a prospective source population. Assuming the null hypothesis that they are drawn from the same parent distribution, the two samples are combined together and randomly resampled into two new samples of the same sizes as the originals, but composed of different data points. In an exact permutation test, all possible resamples are used. In cases where this is combinatorially prohibitive, an approximate permutation test is done by performing $\sim$100--1000 random resamples of the data.

In either case, for each resample, the $k$-nearest neighbors statistic is computed again. The distribution of the statistic over the resamples provides an empirical distribution for what its value ``should'' be under the null hypothesis. If the statistic for the true samples is an outlier in this distribution, we can conclude that they come from statistically different populations. 

An additional challenge of this method is determining the value of $k$ to be used in the $k$-nearest neighbors statistic. Some values of $k$ may indicate a statistically significant result and not others, and each $k$ in some respects represents a different statistic, but they are also correlated with one another because any two statistics $T_{n,k}$ and $T_{n,k+j}$ share the first $k$ neighbors with each other.

This suggests that there should be an optimal value of $k$ for any given test, and we further expect the optimal value to be intermediate between 1 and the total number of points, $n$. At low values of $k$, the variance is high because looking at the single nearest neighbor is much more sensitive to Poisson noise than looking at multiple neighbors. Meanwhile, as $k\rightarrow n$, the entire permutation distribution (including the true value) converges to a single value because every point will be considered as a neighbor to every other. In both cases, the true value is less likely to be an outlier to the distribution, even though it might be for intermediate values of $k$. Broadly, in our tests, the optimal value for $k$ value seems to be near $k\sim\sqrt{n}$.

If the true value of the $k$-nearest neighbors statistic is an outlier of $>2\sigma$ (or alternatively, $p<0.05$), and this is replicated over a range of consecutive $k$, then we can regard it as statistically significant. While a $2\sigma$ anomaly is not considered sufficient for a conclusive detection, it does indicate a statistically significant result to favor or disfavor a particular source population for the sample and therefore merits further investigation.

\section{Results}
\label{sec:results}

Using these statistical methods, we searched for statistically consistent source populations for the host stars of each of the \three categories in our classification scheme, as well as our special dynamical classes. For each category, we considered four possible source populations: all host stars with jupiters, all host stars without jupiters, host stars of multiplanet systems with jupiters, and host stars of multiplanet systems without jupiters. And for each case, we tested on the order of $\sim$30 values of $k$ spaced logarithmically from 1 to $n$. The results of these comparisons are listed in Table \ref{tab:compare}.

\begin{table}[htb]
    \begin{tabular}{ l | l | l | l | l }
    \multicolumn{5}{c}{Consistency of Host Stars of Dynamical Populations of Exoplanet Systems} \\
    \multicolumn{5}{c}{with Potential Source Populations} \\
    \hline
    Population & All & All & Multiplanet & Multiplanet \\
    of Interest & Jupiter-Free Hosts & Jupiter Hosts & Jupiter-Free Hosts & Jupiter Hosts \\
    \hline
    CPP & NO & NO & YES & NO \\
    GPP & Marginal & NO & YES & NO \\
    \hline
    WJ  & NO & Marginal & YES & YES \\
    CWJ & Marginal & YES & Marginal & YES \\
    GWJ & YES & YES &  YES & YES \\
    \hline
    Outer Planets & NO & YES & NO & YES \\
    \hline
    All HJs & NO & NO & NO & NO \\
    HJ Multis & NO & Marginal & NO & YES \\
    HJ+Inner & NO & YES & Marginal & Marginal \\
    HJ+Outer & NO & Marginal & NO & YES \\
    \hline
    IMR & YES & YES & YES & YES \\
    IMR w/o Jupiter & YES & YES & YES & YES \\
    IMR w/Jupiter & NO & YES & YES & YES \\
    \hline
    Super-Puffs ($<30\ M_\oplus$) & NO & YES & Marginal & YES \\
    Super-Puffs ($<60\ M_\oplus$) & Marginal & NO & Marginal & Marginal \\
    Super-Puffs ($<100\ M_\oplus$) & NO & Marginal & NO & YES \\
    \hline
    All USPs & Marginal & NO & Marginal & Marginal \\
    USP Multis & YES & NO & YES & YES \\
    \hline
    \end{tabular}
    \caption{Results of $k$-nearest neighbors permutation tests on host star populations over a range of $k$ from 1 to $>$100 (depending on sample size). Each test compares host stars of a class of exoplanetary systems of interest with those of a potential source population. Tests that remain unambiguously within the $2\sigma$ limits of the permutation distribution are held to be statistically consistent (``YES''). Tests that are clear $>2\sigma$ outliers are inconsistent with the source population and are considered statistically disfavored (``NO''). Tests that fall near the $2\sigma$ limit or exceed it for only one or two values of $k$ are considered to be marginal.}
    \label{tab:compare}
\end{table}

In general, we consider $k$-nearest neighbors statistics that are within the $2\sigma$ bounds ($p\gtrsim0.05$) to be evidence for the ``null'' hypothesis that the sample is consistent with the source population. We consider them to be inconsistent (disfavoring the null hypothesis) if the $k$-nearest neighbors statistics reach values unambiguously $>2\sigma$. And we consider it a marginal detection if the maximum value is near $2\sigma$ or exceeds it for only one or two values of $k$ tested.

Very few of our results in this work exceed $5\sigma$ in significance,\footnote{For individual values of $k$. Consistent values $>2\sigma$ over a range of $k$ should provide more robust results. However, computing their exact significance given the strong correlations between values of $k$ is beyond the scope of this paper.} and it is possible that future discoveries could show some of these categories of exoplanet systems to be consistent with a different source population or populations than our analysis does, especially given the small sample sizes involved. However, the positive results highlighted in this work remain statistically significant ($p<0.05$) and for the most part reinforce the larger trend of larger stars with larger planets. Meanwhile, those populations which are outliers, though not yet conclusive, provide modest evidence for potentially unusual formation and dynamical histories that merit further investigation.

We expect multiplanet host stars to show better agreement with our \three populations of interest than the overall population because of their likely higher completeness of planet discoveries. In Paper I, we found that there appears to be a significant gain in completeness between two-planet and three-planet systems. That is, there are significant differences between the statistics of $N=2$ and $N=3$ systems that are not present between $N=3$ and \four systems, which is why we variously chose three planets, or three \textit{inner} planets, to be sufficient ``completeness'' to fully classify a planetary system under our framework. $N=2$ systems lack this degree of completeness, but are still a significant improvement over single-planet systems for our purposes in this paper.

However, for many categories, multiple or even all of the source populations are consistent with the sample of interest. In these cases, we say that the statistic cannot distinguish between the source populations, so we cannot draw any conclusions about the ``true'' distribution. This is likely to occur if the sample size is very small, without enough data points to clearly indicate it as belonging to one population or another. Given these small samples, the $k$-nearest neighbors statistic can show with some confidence what a particular population of exoplanetary systems is \textit{not}, but it may not be so clear what it \textit{is}, and more exoplanet discoveries, especially in under-explored parts of the parameter space, will be needed to draw more specific conclusions.


\subsection{Closely-Spaced Peas-in-a-Pod Systems}

\begin{figure}[!ht]
    \centering
    \includegraphics[width=0.99\textwidth]{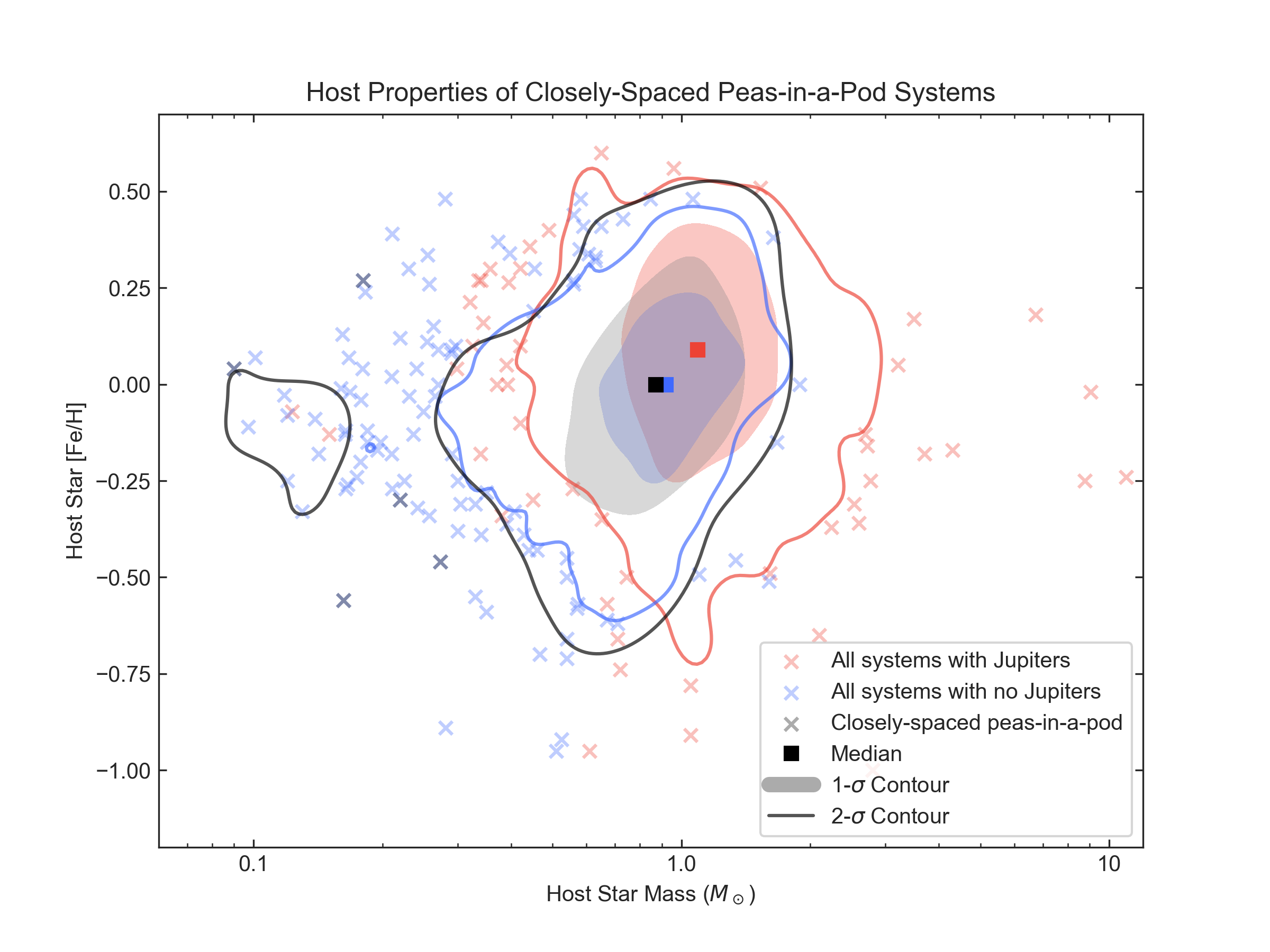}
    \caption{Masses and metallicities of CPP host stars (black) compared with two of our potential source populations: all exoplanet hosts with jupiters (red) and all host stars without jupiters (blue).}
    \label{fig:peapod}
\end{figure}

By far the largest class of \three systems, comprising about three-quarters of the total, are the closely-spaced peas-in-a-pod (CPP) systems: systems with no Jupiters among their inner planets and all of their inner planets dynamically close together, with period ratios $<$5. These systems are numerous enough to plot meaningful density contours that can be compared directly with possible source populations, in addition to our statistical analyses. We plot these density contours in Figure \ref{fig:peapod}. CPP host stars are marked in black compared with all jupiter hosts (red) and jupiter-free host stars (blue).

The contours for CPP host stars appear qualitatively similar to those for jupiter-free hosts more generally, as expected. However, they are wider in the $1\sigma$ contour, and there is also a secondary cluster of CPP hosts at low host masses; and indeed, this is sufficient for our $k$-nearest neighbours statistical test to reject the population of all jupiter-free hosts as a source population. Instead, the only source population that proves to be consistent with the CPP hosts is the \textit{multiplanet} host stars without jupiters. Since this is even closer to our definition for CPP systems, this is the expected result and a good sanity check for our analysis, essentially stating that host stars for $N=2$ and \three systems without jupiters are statistically similar.

\subsection{Other \three Systems}

\begin{figure}[!ht]
    \centering
    \includegraphics[width=0.99\textwidth]{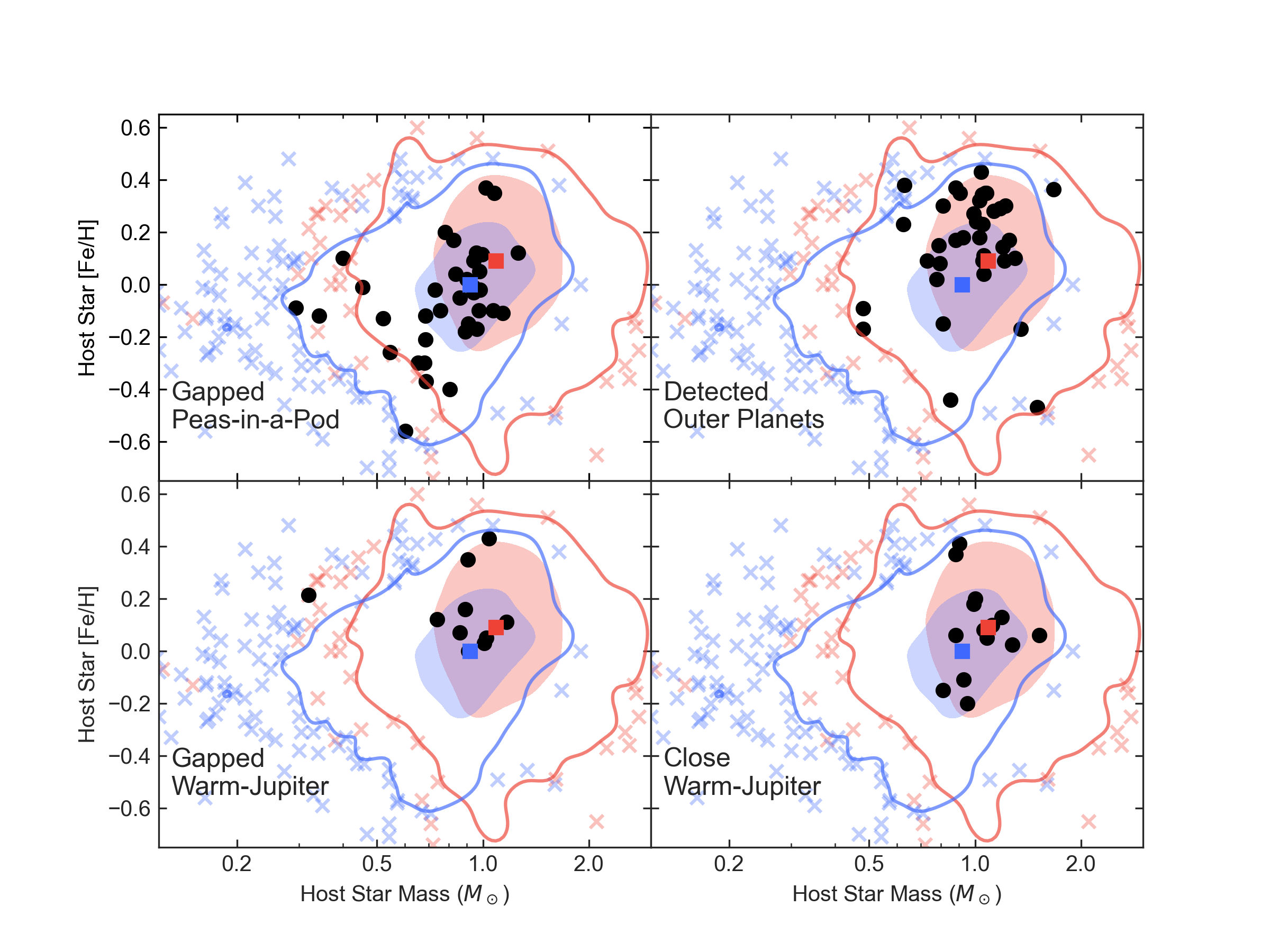}
    \caption{Comparison of host star masses and metallicities of \three categories of exoplanet systems (black points), compared with two of our potential source populations: all exoplanet hosts with jupiters (red) and all host stars without jupiters (blue).}
    \label{fig:classes}
\end{figure}

The other classes of \three systems are too small to plot meaningful density contours. However, for comparison, we also plot their host star properties as scatter plots in Figure \ref{fig:classes} against the same two potential source populations as Figure \ref{fig:peapod}.

The data points for GPP hosts are distributed more similarly to the jupiter-free hosts than the jupiter hosts, and our statistical tests with all four potential source populations give similar results to the CPP hosts. In this case, GPP host stars show marginal agreement with the population of all jupiter-free hosts as well, but the only \textit{clear} agreement is with the population of \textit{multiplanet} jupiter-free host stars, the same as the CPP systems. This is also a good sanity check for the statistics of these smaller samples. 


Warm jupiter (WJ) systems (specifically, those with \three inner planets in our classification) can be divided into closely-spaced (CWJ) and gapped (GWJ) systems, much like the PP systems. Both of these classes appear qualitatively consistent with the larger class of jupiter hosts in Figure \ref{fig:classes}. However, both classes are so small that it is difficult to conclude this for certain. To improve the significance of our result, we tested these classes with our $k$-nearest neighbors statistic both separately and together, but these tests still did not yield statistically significant results, with only one of the 12 clearly \textit{rejecting} a potential source population. Thus, we are not able to conclusively identify the WJ host population with the larger jupiter host population.

\subsection{Systems with Other Dynamical Features of Interest}
\label{sec:special}

\begin{figure}[!ht]
    \centering
    \includegraphics[width=0.99\textwidth]{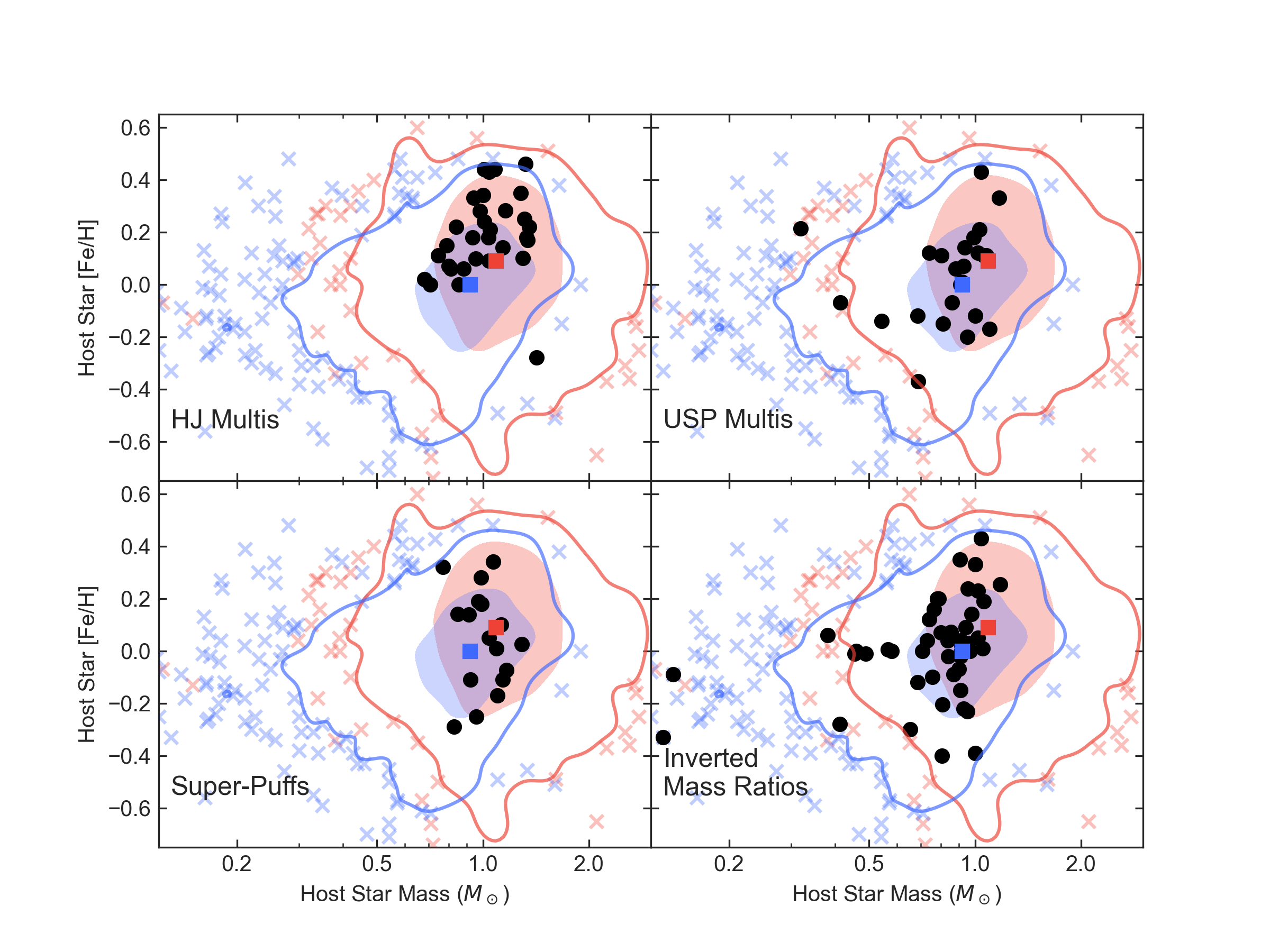}
    \caption{Comparison of host star masses and metallicities of categories of exoplanet systems with unusual dynamical features (black points), compared with two of our potential source populations: all exoplanet hosts with jupiters (red) and all host stars without jupiters (blue).}
    \label{fig:special}
\end{figure}

All of the samples with \three planets in our data set may be expected to resemble the larger population of exoplanet systems as a whole because they are simply a breakdown of that population at higher multiplicities. However, in Paper I, we also highlighted ``other dynamical features'' -- subpopulations of systems that are notable for being unusual in ways that our classification does not otherwise capture. If the properties of a host star have an effect on its planets or on planet formation beyond mere planet size, such effects would be expected to appear more in these populations. We again plot the host star properties of each of these classes as scatter plots against the potential source populations of all jupiter hosts (red) and all jupiter-free hosts (blue) in Figure \ref{fig:special}.

For several of these classes our statistical tests are are not able to shed any light on their host stars. Our tests do not favor any host star population over another for any variation of the ``strongly-inverted mass ratios,'' and we obtain only marginal results for ultra-short period planets. However our results for hot jupiters and super-puffs are more significant.

\subsubsection{Hot Jupiter Multiplanet Systems}
\label{sec:hj}

Beyond the simple categories for dynamical features, multiplanet hot jupiter (HJ) systems have an additional complication in that the population effectively includes two very different architectures. Of the 31 HJ multis in our data set, 22 of them have one or more outer, cold jupiter companions. However, their abundance suggests that they may be statistically normal hot jupiter systems, equivalent to the much larger number isolated hot jupiters that have been discovered. Approximately 3\% of all known hot jupiters have these outer companions, which is the same fraction of outer planets detected for peas-in-a-pod systems. 

In contrast, 10 HJ systems include a hot jupiter together with one or more small, nearby companions.\footnote{There is an overlap of two systems between the two groups, WASP-47 and WASP-132; and one system, WASP-148, is a HJ-WJ pair that does not fit either group.} These systems are much more clearly dynamical outliers from the rest of the HJ population, so we might expect that their host stars could have a different statistical distribution.


We ran $k$-nearest neighbors tests on both the combined HJ multi population and each of the two subsamples. The results of these tests support our hypothesis, but only weakly. All of our tests either reject, or in one case only marginally accept, jupiter-free hosts as a source population, as expected. For jupiter hosts, the host stars of HJs with outer companions show clear agreement with multiplanet jupiter hosts and only marginal agreement with all jupiter hosts, while for host stars of HJs with small \textit{inner} companions, the reverse was true, providing further evidence for HJs with inner companions as dynamical outliers, which may correlate with differences in their host stars.

We note that hot jupiters are themselves a distinct dynamical category, independent of multiplicity. While we focused on the features of multiplanet HJ systems in Paper I, all hot jupiters can be considered members of this category. Therefore, for comparison, we also computed $k$-nearest neighbors statistics for the population of all hot jupiter hosts. However, in this case, \textit{none} of our potential source populations were statistically consistent. It is conceivable that there is a degree of overfitting here, as the much larger sample size (nearly 700), is sensitive to much smaller differences in the distributions. However, on further testing, the HJ hosts proved to be most similar to host stars with \textit{multiple} jupiters: slightly higher in mass and metallicity on average than other giant planet hosts.




\subsubsection{Super-Puffs}
\label{sec:sp}

Super-puffs are low-mass planets with anomalously low densities, beyond what can be easily explained by young ages or radius inflation. While there is no agreed-upon definition for super-puffs, in Paper I, we adopted a definition of planets with $M<30\,M_\oplus$ and $\rho<0.3$ g cm$^{-3}$.\footnote{The super-puffs were the class of systems proportionally most affected by the updated exoplanet census, with one planet added, Kepler-90g, and two planets removed due to updated mass or radius measurements: Kepler-105b and TOI-1836b.} Super-puffs, in addition to being physically anomalous, prove to be very unusual dynamically in that they occur almost exclusively in multi-planet systems.\footnote{This may be partly because many super-puffs have masses measured by transit timing variations, which can be detected only in multiplanet systems. However, the number of radial velocity detections of super-puffs in multiplanet systems is still large enough to be anomalous.} Of the 18 systems in our dataset containing super-puffs under our definition, 16 of them are multi-planet systems. (Most of their companions are \textit{not} super-puffs themselves.)

In addition to the anomalous number of multiplanet systems, our $k$-nearest neighbors tests for super-puff host stars yielded another surprising result: 
super-puff host stars appear to favor \textit{jupiter} hosts as a source population at a statistically significant level, rejecting the jupiter-free host population at $\sim2.5\sigma$ ($p\approx0.01$). This is despite the fact that only 5 out of 18 super-puff systems have jovian companions. This result is not quite as strong when considering multiplanet host stars as a source population, as indicated in Table \ref{tab:compare}, but it is still unexpected. A statistical distribution of host stars matching those of jupiter hosts suggests that super-puffs have more massive and higher-metallicity host stars compared with otherwise-similar systems without super-puffs. This result combined with the even stronger correlation with multiplanet systems may suggest particularly unusual formation histories for these planets.

Because the definition used for super-puffs can vary (also resulting in a variation in the \textit{number} of super-puffs by a factor of a few, we also consider populations with higher mass limits to capture the inflated sub-saturn regime, which may constitute an extension of it. In Table \ref{tab:compare}, we list $k$-nearest neighbors tests for ``super-puff'' classes with a maximum mass of 60 $M_\oplus$ and 100 $M_\oplus$. In the 60 $M_\oplus$ case, there are no strong matches to the super-puff population, although its remain qualitatively similar to the other cases in mass-radius space. This may be due to random variation given the small number of systems, or it may be evidence that the lower-mass super-puffs are truly statistically distinct from the inflated sub-saturns. Meanwhile, the 100 $M_\oplus$ case is an even stronger match to the multi-planet jupiter hosts in particular. This may be because the planets in the 60-100 $M_\oplus$ range are more representative of our jupiter sample to begin with. However, we note that even under the broadest definition, the super-puff sample suffers from small number statistics, and more data will be needed to reach more definitive conclusions.

\section{Discussion}
\label{sec:discussion}

The specific dynamical architectures of exoplanetary systems show little correlation with the properties of their host stars, with few exceptions. In some cases, the lack of statistically significant results is likely due to the small sample sizes for many dynamical classes, but even these small samples rule out large deviations from the general host star population.

By far the strongest correlation is found between host star properties and planet mass, independent of dynamics, following the well-known result that larger stars tend to have larger planets \citep{Laughlin2004}, and more metal-rich stars tend to have larger planets \citep{Fischer2005}. Our results in this paper support that conclusion and extend the observed trends to higher total planet mass. Specifically, we find that stars with \textit{multiple} jupiters tend to have higher metallicities than stars where only one jovian planet has been detected.

Likewise, our finding that the upper limit of planet masses is broadly consistent with models of the stability of protoplanetary disks should be a robust result, given the high detectability of such large planets. Taken together, these two results, relying in large part of the distribution of the largest planets, are our strongest findings, confirming and extending previous studies.

Beyond the aforementioned trends, we see few distinguishing patterns in any subpopulation. The differences between most subpopulations can be adequately explained by formation models that predict correlations with overall sizes and numbers of planets, without the need for any more complex dependence on host star properties. This finding suggests that stellar properties do not have a large effect on the planet formation process, nor (unsurprisingly) on the system's dynamical history.

We note that these results are limited by the fact that the current census of exoplanets includes few mid-to-late M-dwarfs. These low-mass stars are known to be distinct in the exoplanet population, with an even lower occurrence of giant planets \citep[see, e.g.,][]{GEMS2}, but the sample size remains too small to reveal significant population-level trends. It is possible that a larger sample of planets orbiting low-mass stars will indicate larger dynamical or formation-related differences from planets of larger stars.

Current classification efforts are limited by observational biases in the observed exoplanet population. For example, while the paucity of giant planets around low-mass stars appears to be robust, the reverse, the relative lack of small planets around higher-mass stars, is much more affected by the low detectability of such planets. Nonetheless, our comparison of samples within the observed exoplanet population should help mitigate this problem, as each sample will have similar biases.

\subsection{Super-Puffs}

While most exoplanet architectures considered in this work did not show statistically significant correlations with host star properties, systems containing super-puffs represent a notable exception. We find that super-puffs are demographically similar to Jupiters in multi-planet systems (in that they orbit higher-mass and higher-metallicity stars on average), despite having Neptune/sub-Neptune masses. They also occur disproportionately in multi-planet systems. This correlation presents a puzzle which is compounded by the fact that super-puff planets, as a population, appear to be diverse compared with other sub-populations in the exoplanet sample. 

Super-puffs appear to attain their remarkably low densities through different underlying mechanisms, which vary from planet to planet. For example, short-period super-puffs may have radii inflated by high levels of stellar irradiation \citep{Guillot2002, Laughlin2011, Thorngren2018}, a process that depends on stellar mass. Dynamical effects may also increase the radii of these planets via mechanisms such as Ohmic dissipation \citep[as in WASP-107;][]{Batygin2025} or tidal inflation \citep{Millholland2019}, in a way that is also stellar-mass dependent.

Super-puffs young enough to retain some heat of formation \citep[such as the three super-puffs in the Kepler-51 system;][]{Steffen2013, Masuda2014}, while not sufficient to inflate their radii to the observed degree directly, may have sufficient heat to drive outflows of small grains produced by photochemical hazes \citep{Gao2020,Ohno2021}. On the other hand, these mechanisms do not adequately explain the observed radii of old, cold, long-period super-puffs. Instead, circumplanetary rings have been proposed to explain the radii of planets such as HIP 41378 f and Kepler-79d \citep{Piro2020, Lu2025}. Although no direct evidence for rings around super-puffs has yet been found \citep{Umetani2025}, various theoretical and observational analyses have thus far supported (or at least not rejected) the feasibility of rings around these planets \citep{Alam2022, Ohno2022,  Belkovski2022}.

Our observed correlation between super-puffs and multiplanet systems hosting Jupiters suggests that processes more prevalent in such environments may influence the inflated radii of these planets. For instance, collisions between bodies could heat planetary envelopes, temporarily inflating planetary radii and lowering inferred densities \citep{Anderson2012, Biersteker2019}. Additionally, dynamical interactions in multi-body (planet/moon) systems may facilitate the formation of circumplanetary rings \citep{Saillenfest2023}. Both mechanisms are more likely to occur in more complex multi-body architectures.

A final explanation for the diversity of the super-puff sample could also be misclassifications (as indeed was the case for Kepler-105b and TOI-1836b). Some candidate super-puffs have poorly determined masses \citep[for example, Kepler-359d, which has masses obtained via TTVs and mass uncertainties above 90\%;][]{Hadden2017} and as a result may be miscategorized as super-puffs. These diverse explanations for super-puff radii highlight the fact that super-puffs are an extremely non-homogeneous population.

\subsection{Hot Jupiters}

In Paper I, we remarked upon the small subset of hot jupiters with small interior companions, but we did not identify them as a potential additional class of planetary systems at that time. Hot jupiters are a distinct dynamical class in themselves, and those with nearby companions, if they have at least three inner planets as WASP-47 and K2-19 do, also qualify for our definition of ``warm jupiter systems.'' However, the category of ``hot jupiters with close friends'' has been recognized as unusual and potentially important \citep[see, e.g.,][]{Rodriguez2025}, and they may well be deserving of their own class in our system (see also \citealt{Wu2023}). 

The results of this paper provide modest evidence in support of this idea. Host stars of hot jupiters with inner companions show better statistical agreement with jupiter hosts as a whole, as opposed to multiplanet hosts, unlike the other subcategories we tested. While these systems are themselves multiplanet systems, it does suggest a statistically different population from both the host stars of HJs with outer companions, and host stars of HJs as a whole. This evidence is not yet sufficient to definitively identify such a trend, but it underscores the need to find more ``hot jupiters with friends'' to better understand this population.

\subsection{Planet-Star Mass Ratios}

The data show a modest correlation of total planet mass with the mass of the host star (see Figure \ref{fig:starprop}).  In addition, the total planet mass shows a well-defined upper limit, corresponding to a total mass of $\sim10^4\,M_\oplus$ and a mass ratio of $M/M_\ast\sim0.03$. These findings are broadly consistent with expectations from star formation theory. Stability considerations in circumstellar disks suggest that the maximum disk mass, and hence the maximum mass available for the formation of planets, corresponds to a mass ratio $M_d/M_\ast\sim0.1$ (see \citealt{adams1989,shu1990}, and many subsequent studies). Star/disk systems with mass ratios above this threshold are subject to robust gravitational instabilities that act to redistribute mass toward a more stable configuration. The time scale for the growth of these instabilities is determined by the orbit time at the outer disk edge, typically $\sim1000$ yr, so we expect planet-forming disks to have mass ratios below this threshold. The moderate values of the total planet mass found here are thus consistent with this expectation. Moreover, the absolute value of the total planet mass falls just below the benchmark value for the Minimum Mass Solar Nebula \citep{hayashi1981}, again consistent with expectations.  

Although the mass ratios described above are consistent with expectations, note that the current catalog of planetary companions is not complete. If sufficient numbers of large planets are present, but not yet detected, then the upper envelope could be higher than the quoted value of $M/M_\ast\sim0.03$. Because this value corresponds to 30 Jovian masses for a 1 $M_\odot$ host star, and the systems are unlikely to harbor $\sim30$ unseen Jupiters, the putative unseen bodies would have to much larger in order to substantially change the above interpretation. In addition, the theoretical maximum ($M_d/M_\ast\sim0.1$) for disks is three times the currently inferred limit. Nonetheless, it remains possible for brown dwarfs to form out of self-gravitating disks, so that unseen brown dwarfs could increase this limit. The preferred mass scale for forming secondary bodies in a disk is estimated to be of order $M_S\sim10M_J$ \citep{xu2025,adams2025}, so that a large number ($\sim7$) of brown dwarfs would be necessary to violate theoretical expectations. Finally, we note that (as yet unseen) stellar companions could increase the mass ratios beyond the $M_d/M_\ast\sim0.1$ threshold. However, stellar companions are unlikely to form out of pre-existing, stable, centrifugally supported disks, but rather result from earlier fragmentation events (see the aforementioned papers as well as \citealt{kratter2010,kratter2016,Offner2023}).

\subsection{Interpreting the Absence of Trends in Other Categories}

In the past, numerous studies have shown that planetary properties, such as size \citep{Buchhave2012, Buchhave2014} and occurrence rates \citep{Fischer2005, Johnson2010occur, Reffert2015, Ghezzi2018, Gan2025}, are correlated with host star properties, particularly stellar mass and metallicity. These trends have helped inform theories of planet formation and evolution, particularly in the context of core accretion and disk-driven growth \citep{ppvii}. However, most of this prior work has focused on individual planet properties rather than the architectures of full planetary systems.

When considering a variety of planetary architectures, for most populations, we find no strong evidence that architectures vary significantly with stellar mass or metallicity. While stellar properties may shape whether planets form in the first place \citep{Pollack1996}, we do not find evidence that they exert a similarly strong influence on how most classes of planetary systems are dynamically assembled.

This lack of a strong correlation implies that the dynamical processes responsible for sculpting system architectures, such as planet-planet interactions, migration, and instabilities, are either largely independent of the host star’s mass and composition, or any existing correlations are too subtle to be detected in our current dataset. Put differently, from current evidence, the mechanisms shaping planetary system architectures seem to be primarily dynamical in nature and not directly tied to stellar properties.

That said, there remain populations and regimes where the current data are insufficient to disentangle underlying trends between planetary architectures and stellar properties. In some of these cases, we cannot even detect the \textit{known} correlation between larger stars and larger planets, pointing to observational limitations in the observed planet population preventing the detection of statistically significant differences. These ambiguities could be resolved in the future with additional data.

On a broader scale, this work reaffirms an accepted finding in exoplanet demographics: for single-planet systems, it is generally easier to form a Jupiter-mass planet around a more massive star, and correspondingly harder around a lower-mass star (Figure \ref{fig:nsize}). This trend holds broadly across the $M_\ast=0.4 – 2.0~M_\odot$ range and likely represents a first-order result that aligns well with theoretical expectations \citep{Laughlin2004}. This work generalizes the trend by showing that the total mass in planets also correlates with host star mass. There are, of course, exceptions where large planets orbiting low-mass stars have been observed \citep[e.g.,][]{Johnson2010, Kanodia2024}, but these remain outliers rather than the norm. For Jupiter-mass planets in tight orbits, a similar trend holds with stellar metallicity \citep{Fischer2005}. Additional correlations between planetary system characteristics and stellar host properties await the availability of more data. 

\vspace{12pt}

\section*{Acknowledgments}

ARH acknowledges support by NASA under award number 80GSFC24M0006 through the CRESST II cooperative agreement, as well as support from the GSFC Exoplanet Spectroscopy Technologies (ExoSpec).

FCA acknowledges support from the Leinweber Institute for Theoretical Physics (LITP) at the University of Michigan, as well as NSF Grant No. 2508843.

This work was performed in part by members of the Virtual Planetary Laboratory Team, a member of the NASA Nexus for Exoplanet System Science, funded via NASA Astrobiology Program Grant No. 80NSSC18K0829.

We thank Noah Tuchow for his advice for developing the statistical analysis used in this paper. We also thank Chris Stark and Dan Fabrycky for helpful conversations.

We thank the anonymous referee for their assistance in improving the quality of this paper.

This research has made use of the NASA Exoplanet Archive, which is operated by the California Institute of Technology, under contract with the National Aeronautics and Space Administration under the Exoplanet Exploration Program.

This paper makes use of data from the first public release of the WASP data \citep{Butters10} as provided by the WASP consortium and services at the NASA Exoplanet Archive, which is operated by the California Institute of Technology, under contract with the National Aeronautics and Space Administration under the Exoplanet Exploration Program.

This paper makes use of data from the UKIRT microlensing surveys \citep{Shvartzvald17} provided by the UKIRT Microlensing Team and services at the NASA Exoplanet Archive, which is operated by the California Institute of Technology, under contract with the National Aeronautics and Space Administration under the Exoplanet Exploration Program.

This paper makes use of data from the KELT survey, which are made available to the community through the Exoplanet Archive on behalf of the KELT project team.

This paper makes use of data obtained by the MOA collaboration with the 1.8-metre MOA-II telescope at the University of Canterbury Mount John Observatory, Lake Tekapo, New Zealand. The MOA collaboration is supported by JSPS KAKENHI grant and the Royal Society of New Zealand Marsden Fund. These data are made available using services at the NASA Exoplanet Archive, which is operated by the California Institute of Technology, under contract with the National Aeronautics and Space Administration under the Exoplanet Exploration Program.

\software{matplotlib \citep{Hunter:2007},
pandas \citep{mckinney-proc-scipy-2010},
scipy \citep{2020SciPy-NMeth}, scikit-learn \citep{scikit-learn}, seaborn \citep{Waskom2021}}

\facilities{Exoplanet Archive \citep{Christiansen2025}}

\bibliography{refs}
\bibliographystyle{aasjournal}   

\end{document}